\documentclass[12pt]{article}
\usepackage{amsfonts,latexsym,eucal,color,graphicx,epsfig,amssymb,amsmath}

\def\ben{\begin{equation}}
\def\een{\end{equation}}
\def\half{{1 \over 2}}
\def\bea{\begin{eqnarray}}
\def\eea{\end{eqnarray}}
\def\p{\partial}

\newcommand{\pd}{\partial}

\begin{document}
\begin{titlepage}
\begin{flushright}
	{\small DAMTP-2009-43}\\
	{\small WU-AP/300/09}\\
	{\small arXiv:0906.0264 [hep-th]}\\
\end{flushright}
\vspace{1.5cm}
\begin{center}
	{\Large {\bf The Bernstein Conjecture, Minimal Cones,\\ and Critical Dimensions}}\\
\vspace{1.2cm}
	Gary W.~Gibbons$^{\dagger}$,
	Kei-ichi Maeda$^{\ast}$,
	and
	Umpei Miyamoto$^{\sharp}$
\\
\vspace{.6cm}
	{\it
		$^{\dagger}$DAMTP, Centre for Mathematical Sciences, Cambridge University,\\ Wilberforce Road,
		Cambridge CB3 OWA, UK \\
		$^{\ast}$Department of Physics, Waseda University,\\ Okubo 3-4-1, Tokyo 169-8555, Japan \\
		$^{\sharp}$Racah Institute of Physics, Hebrew University of Jerusalem,\\ Givat Ram, Jerusalem 91904, Israel\\
	}
\vspace{.6cm}
	\small{\tt{
		G.W.Gibbons@damtp.cam.ac.uk,
		maeda@waseda.jp,
		umpei@phys.huji.ac.il
	}}\\
\vspace{1.2cm}
\end{center}
\begin{abstract}
Minimal surfaces and domain walls play important roles in various contexts of spacetime physics as well as material science. In this paper, we first review the Bernstein conjecture, which asserts that a plane is the only globally well defined solution of the minimal surface equation which is a single valued graph over a hyperplane in flat spaces, and its failure in higher dimensions. Then, we review how minimal cones in four- and higher-dimensional spacetimes, which are curved and even singular at the apex,  may be used to provide counterexamples to the conjecture. The physical implications of these counterexamples in curved spacetimes are discussed from various points of view, ranging from classical general relativity, brane physics, and holographic models of fundamental interactions.
\end{abstract}

\vspace{1.2cm}
\begin{flushleft}
	{\small June 1, 2009}
\end{flushleft}

\end{titlepage}

\tableofcontents

\section{Introduction}
\label{sec:intro}

Minimal surfaces\footnote{Newcomers to the subject are warned that 
according to the current mathematical usage,
which is sanctioned by a long standing 
but nevertheless illogical and confusing tradition, the words \lq\lq minimal
surface\rq\rq  are used to mean a $p$-dimensional surface, {\it i.e.}, a $p$-brane,
whose first variation
of the $p$-volume functional vanishes. There is no   
implication about the
second variation, {\it i.e.}, the Hessian of the  $p$-volume functional. 
Therefore it may or may not be the case that a minimal surface is a 
 (local or global) \lq\lq minimizer\rq\rq  of the $p$-volume functional. Extremal surface would be a better name.
Note that when we speak of a variational
principle for non-compact surfaces we mean, as is customary in the
mathematical literature, against {\it compactly supported variations}.
In other words, in any calculation we only consider an integral of a compact
subset of the non-compact surface and test its second variation.  
A further potential 
source of confusion is that
a minimal surface is often referred to as being \lq\lq stable\rq\rq if 
the Hessian    of the $p$-volume functional is positive definite. 
Stability in this sense  may or may not
coincide with a dynamical notion of 
stability, depending on the physical context.} in Euclidean space ${\Bbb E} ^3$  have been extensively studied since the pioneering work of Thomas Young and of Laplace.
In Monge, or non-parametric,  gauge the surface is  specified by   the height function $z=z(x,y)$ 
above some plane. The non-parametric minimal surface equation governing the function  $z(x,y) $ is  
\ben
	\p_x \left({ z_x \over \sqrt{ 1 + z_x^2 + z_y^2 } } \right)
	+
	\p_y \left({ z_y \over \sqrt{ 1 + z_x^2 + z_y^2 } } \right)
	= 0\,.
\label{min}
\een

A famous result of Bernstein asserts that the only single valued solution of Eq.~(\ref{min}) defined for all $(x,y) \in{\Bbb R} ^2$
is a plane. It may also be shown that the planar solution is a minimizer of the area functional
among  compactly supported variations of the surface. In terms of brane theory, this means that the  \lq \lq classical ground state\rq \rq , {\it i.e.}, the static minimum of the energy functional
for a  membrane  in three dimensional Euclidean space ${\Bbb E} ^3 $,  
which may be thought of as a static configuration 
in   4-dimensional Minkowski spacetime ${\Bbb E} ^{3,1} $,  is 
smooth and indeed planar.    
From the world volume point of view the  classical ground state of the membrane
preserves (2+1)-dimensional  Poincar\'e invariance and may be thought of as 
a copy of  (2+1)-dimensional Minkowski spacetime ${\Bbb E} ^{2,1}$.

It is natural to conjecture that Bernstein's theorem remains  valid
for a 
minimal $p$-dimensional hypersurface  in  
$(p+1)$-dimensional Euclidean space ${\Bbb E}^{p+1}$.
In other words, the classical ground state of a {\it $p$-brane}
in $(p+1+1)$-dimensional Minkowski spacetime  ${\Bbb E} ^{p+1,1}$ 
should be flat and invariant under the action of the 
$(p+1)$-dimensional Poincar\'e group $E(p,1)$.  Remarkably, 
although true for $p\le 7$  it fails for $p+1=9$~\cite{Bombieri}.
In other words, the classical ground state of an 8-brane
in 10-dimensional Minkowski spacetime spontaneously breaks
(8+1)-dimensional Poincar\'e invariance.
The proof~\cite{Bombieri} rests on the fact 
that in ${\Bbb E}^8$ and above, a minimal hypersurface which is a
minimizer of the $p$-volume  functional among  compactly 
supported variations  need not be smooth.
There are rather explicit counterexamples
called {\it minimal cones}. Their existence leads to the conclusion
that Bernstein's theorem fails in ${\Bbb E}^9$~\cite{Bombieri} \footnote{
We emphasize for clarity that Bombieri et alia's result uses the failure
of regularity of seven dimensional minimal hypersurfaces in 8 
ambient dimensions to establish the non-uniqueness of
({\it i.e.}, the existence of non-flat) 
minimal hypersurfaces of co-dimension one which may be given by a 
single valued height function
defined for all of ${\Bbb R } ^ 8 $ in 9 ambient dimensions. The argument,
which depends on scaling properties of the equations, is given in detail
in the cited references.}. 
 
As far as we aware, there has been very little discussion of the 
significance of this fact in the M/String theory literature.
The breakdown of regularity of minimal hypersurfaces of flat space 
extends to minimal hypersurfaces 
of curved Riemannian manifolds and this has consequences for proofs of the 
positive energy theorem  
of general relativity which make essential use of minimal surfaces as a 
technical tool~\cite{Schoen1,Bray,Schwartz:2007gj}.      
It seems worthwhile therefore to examine the behavior of minimal
surfaces in higher dimensions and in curved
spaces in some explicit detail in order to  understand  
better the situation and its possible physical
implications. In particular, it is interesting to see whether the 
existence of various critical dimensions which has been noted in related contexts is of a 
universal nature and related to the the breakdown of Bernstein's theorem and the existence of minimal cones.

To make  progress it is helpful to assume that the relevant surfaces have 
sufficient symmetries with which the problem may be reduced to one involving ordinary 
differential equations in an appropriate quotient space $X$, a 
ploy known to mathematicians as {\it equivariant variational theory}.
Typically the brane equations of motion reduce
to finding geodesics in $X$ with respect to a suitable metric $g$ on $X$, induced by the $p$-volume functional.
The $p$-brane  will be $p$-volume minimizing if the corresponding geodesic $\gamma$ is length minimizing.
A necessary condition that a geodesic joining points
$a$ and $b$ be length minimizing is that $\gamma$  
contains no points between $a$ and $b$ conjugate to either.
The existence of such conjugate points is governed by the Jacobi or
geodesic deviation equation, solutions of which
depend on the curvature of $X$. In the case that $X$ is 2-dimensional, it is the sign
of the Gauss curvature $K$ which is important. 
If for example $K$ is negative in the vicinity of $\gamma$, then it
can contain no conjugate points and hence must be locally length minimizing.
In the cases we shall consider the Gauss curvature is actually  positive
and a more detailed examination is required. One might have thought
that positive Gauss  curvature  would lead to a  second variation
or Hessian of indefinite sign. However, the situation is more subtle
since the effective metric governing the variational
principle is incomplete and  becomes singular near a conical point 
and compensatory terms can arise which in low dimension
render the Hessian positive definite.

The simplest example of this situation is when a 
Lie group $G$ acts isometrically on a 
$(p+1)$-dimensional Riemannian manifold $\{\Sigma, h\} $, thought of as 
``space'' with orbits which are $(p-1)$-dimensional. The $X= \Sigma/G$ is (locally) 2-dimensional 
and a curve in $X$ may be thought of as the projection under the action of $G$ of 
a $p$-dimensional submanifold $S$ of $\Sigma$.
The basic example of this setup is when $p=2k+1$, $\Sigma= {\Bbb E} ^{2k+2}$ and $G= \mathrm{SO}(k+1) \times \mathrm{SO}(k+1)$ with
the standard action on ${\Bbb E} ^{2k+2} = {\Bbb E} ^{k+1} \times {\Bbb E} ^{k+1}$ with flat metric  
\ben
	h = dx^2 + x^2 d \Omega^2  _k + dy^2 + y^2 d \Omega_k ^2\,, 
\een   
where $d \Omega _k^2$ is the standard round metric on $S^k$. The induced metric $g$ is
\ben
	g = (xy)^{2k} \bigl(dx^2 + d y^2 \bigr)\,.
\een   
The orbit of the straight line $x=y$ under the action of $\mathrm{SO}(k+1) \times
\mathrm{SO}(k+1)$  is a $(2k+1)$-dimensional  minimal cone
with a singularity at the origin, $x=y=0$. A study of the second variation shows that this singular cone is
$(2k+1)$-volume minimizing as long as $k\ge 3$.
In the next section we review the existence of minimal cones in higher dimensional flat spaces with more general examples. From the physical point of view, the minimal surface is a thin limit of a domain wall. Thus, also in the next section, we describe the relation between the Bernstein conjecture and some theorems/conjectures asserting the existence of non-planar domain walls.
Then, in Sec.~\ref{sec:BH} we present a general prescription to investigate minimal surfaces in curved background and explicit examples of minimal cones in black hole backgrounds. We show that the critical dimension in the black-hole--brane system investigated by Frolov~\cite{Frolov} is related to the failure of the Bernstein conjecture. In addition, we present a new example of a minimal cone in the black hole background. The final section is devoted to discussions of possibly related systems.

\section{Branes in Flat Backgrounds}
\label{sec:flat}

\subsection{A Primary Example}
\label{sec:primary}

By treating a simple but nevertheless important example, 
we see that the problem of  finding  minimal surfaces  reduces to finding 
geodesics in a low dimensional space. 
Then, the criterion of length minimizing for the geodesic is introduced. One may find more mathematical 
and general introduction in~\cite{Fomenko}. 

Let us consider a $G$-invariant submanifold $S \subset \Sigma = \mathbb{E}^{m+n+2}$ ($m,n \geq 1$) as the configuration of $p$-brane ($p=n+m+1$). For a simple example, we focus on the case of $ G=\mathrm{SO}(m+1) \times \mathrm{SO}(n+1) \subset \mathrm{SO}(m+n+2)$.
The metric of ambient flat space is written in the form
\bea
	h
	=
	dx^2 + x^2 d\Omega_m^2 + dy^2 + y^2 d\Omega_n^2\,.
\eea
A projected metric $\tilde{h}$ on $\Sigma/G$ is defined via a transformation of an inner product,
$ \tilde{h}(u_1,u_2) = h(u_1,u_2) $, where $(u_1,u_2)$ is a pair of vectors tangent to the orbit $G(s)$, $s \in S$.
Thus, in the present case
\bea
	\tilde{h}
	=
	dx^2 + dy^2\,.
\eea
The volume function on $\Sigma/G$, which is the volume of the orbit $G(s)$, is given by
\bea
	v(x,y)
	:=
	\mathrm{Vol}\,\left( G(s) \right)
	=
	\Omega_m x^m \Omega_n  y^n\,,
\eea
where $\Omega_n$ is the volume of unit $n$-sphere. Now, we are ready to define an effective space where the geodesic $\gamma$ is to be found. The metric $g$ of the effective space is defined as
\bea
	g
	=
	d\ell^2
	:=
	v^{2/\lambda}(x,y) \tilde{h}
	=
	\Omega_m^2 \Omega_n^2  x^{2m} y^{2n} ( dx^2 + dy^2 )\,,
\eea
where $\lambda$ is the co-dimension of the $G$-invariant surface $S$.
For the present example, $\lambda$ is given by
\bea
	\lambda
	:=
	\mathrm{dim}\, S - \mathrm{dim}\, G(x)
	=
	(m+n+1) - (m+n)
	=
	1\,.
\eea
Hereafter, we omit the unimportant numerical factor of $g$, $\Omega_m^2 \Omega_n^2$.

The problem to find the minimal surface has been reduced to find the geodesic $\gamma$
with $g=d\ell^2$. The action to be minimized is
\bea
	\ell
	=
	\int x^m y^n \sqrt{ dx^2 + dy^2 }\,.
\eea
If one denotes the geodesic by $y=y(x)$ and varies the action with respect to it, one has
\bea
	xy y^{\prime\prime}
	+
	( myy^\prime - nx ) ( 1+y^{\prime 2} ) =0\,.
\label{eq:EL}
\eea
One can easily see that the following cone solves Eq.~(\ref{eq:EL}),
\bea
	y = \sqrt{\frac{n}{m}}\;x\,.
\label{eq:cone}
\eea

As noted before, one cannot know whether the above cone is indeed a 
minimizer or not until examining the second variation. 
Here, we introduce an alternative criterion to 
examine whether a geodesic is a minimizer or not.
Readers are directed to~\cite{Rauch} for an introduction to this topic. The {\it Jacobi equation} or {\it equation of geodesic deviation} is written as
\bea
	{d ^2 \eta \over d \ell ^2 } +  K \eta = 0\,,
\label{eq:gd}
\eea
where $\eta$ and $K$ are the geodesic deviation and Gauss curvature of metric $g$.
For a general metric of form $g=v(x,y)^{2/\lambda}(dx^2+dy^2)$, the Gauss curvature is given by
$ K = (1/2) ( \mathrm{Ricci \; scalar} ) =- 2\lambda^{-1} v^{-2/\lambda} ( \pd_x^2 + \pd_y^2 ) \ln v $.
For the present case, where $v(x,y)=x^my^n$ and $\lambda = 1$, we have
\bea
	K
	=
	\frac{1}{x^{2m}y^{2n}}
	\left(
		\frac{ m }{ x^2 } + \frac{ n }{ y^2 }
	\right)\,,
\eea
which is positive definite.
One can calculate the Gauss curvature and proper distance along the geodesic~(\ref{eq:cone}),
\bea
	K
	=
	\frac{ 2  m^{n+1} }{ n^n x^{2(m+n+1)} }\,,
\;\;\;\;
	\ell
	=
	\frac{ n^{n/2} (m+n)^{1/2} }{ m^{(n+1)/2} (m+n+1)} x^{m+n+1}\,.
\label{eq:along-cone}
\eea
Combining Eqs.~(\ref{eq:gd}) and (\ref{eq:along-cone}), we have
\bea
	\frac{ d^2 \eta }{ d\ell^2 } + \frac{c}{\ell^2} \eta = 0\,,
\label{eq:Jacobi}
\eea
where
\bea
	c = \frac{ 2(m+n) }{ ( m+n+1 )^2 }\,.
\eea
It is well known that the behavior of solution to this equation changes at $c=1/4$.
That is, Eq.~(\ref{eq:Jacobi}) has a simple power solution
\bea
	\eta = \ell^{\beta_\pm}\,,
\;\;\;
	\beta_\pm
	=
	\frac12 \left( 1 \pm \sqrt{ 1-4c } \right)\,.
\eea
Thus, the geodesic deviation oscillates ({\it i.e.}, there exists a conjugate point of the geodesic) for $ 2 \leq m+n \leq 5 $, while not for $ m+n \geq 6 $ ({\it i.e.}, there exists no conjugate point of the geodesic).
These results, and further work by Bombieri {\it et al.}~imply that the cone (\ref{eq:cone}) as $\mathrm{SO}(m+1) \times \mathrm{SO}(n+1)$-invariant hypersurface $S \subset {\Bbb E}^{m+n+2}$ is a minimizer for $ m+n+2 \geq 8$.

\subsection{Other Symmetry Groups and Higher Co-dimensions}

We have seen the existence of $\mathrm{SO}(m+1) \times \mathrm{SO}(n+1)$-invariant minimal cone above.
In Ref.~\cite{Fomenko} Fomenko gives further examples
and a classification scheme for  minimal cones in ${\Bbb E}^{p+1} $.
The starting points are minimal $(p-1)$-dimensional submanifolds $S = G/H
\subset S^p \subset {\Bbb E}^{p+1} $ of the round $p$-sphere invariant under $G \subset \mathrm{SO}(p+1)$ with
a stabilizer or isotropy group $H \subset G$. 
There are 12 possibilities (p.~100 in~\cite{Fomenko}).  
As before $x$ and $y$ are coordinates on the 2-dimensional quotient space $X= {\Bbb E}^{p+1}/G$. Letting $v(x,y)$ be the volume of the orbits, the effective metric on $X$ is again given by $ g = v^2(x,y) ( dx^2 + dy^2 )$. The minimal cones are given by straight lines in the ($x,y$)-space.\footnote{From the analogous point of view in optics, the function $v(x,y)$ plays the same role as the refractive index in Fermat's principle.}
Reference~\cite{Fomenko} (Theorem 2 on p.~103) states which of the possibilities
listed (Table 1 on p.~102) are actually minimizing. The critical dimension of Euclidean space varies from group to group, but is never less than the 8 found by~\cite{Bombieri}.

Reference~\cite{DaoFomenko} (pp.~146-147) 
contains the material similar to Ref.~\cite{Fomenko}.
The classification theorem is the same as~\cite{Fomenko} 
but the authors also quote the results of Ivanov on $p$-dimensional cones in  ${\Bbb E}^{p+2}$, {\it i.e.},
on minimal surfaces of co-dimension two.

Although we will focus on  co-dimension one 
cones, especially the case of $G=\mathrm{SO}(n+1) \times \mathrm{SO}(m+1)$, 
in the rest of paper, it would be interesting to investigate 
the physical implications of the other  types of minimal cones
mentioned above.

\subsection{The Bernstein Conjecture and Domain Walls}

From a physical point of view, a minimal surface is a mathematical idealization of something with finite thickness. A model which incorporates this is a non-linear Laplace equation of the
form
\ben
	\Delta \phi = V^\prime(\phi)\,,
\;\;\;\;
	\Delta = \sum_{i=1}^{p+1} \frac{ \pd^2 }{ \pd x_i^2 }\,.
\label{eqn}
\een
If $V(\phi)$ has two critical points at $\phi =\pm1$, say, at which
\ben
	V^\prime (\pm 1) = V(\pm1) =0\,,
\een
then a  static domain wall is a solution on ${\Bbb E}^{p+1}$, with $\phi \rightarrow +1 $ as $x_{p+1} \rightarrow +\infty$ and $\phi \rightarrow -1 $ as $x_{p+1} \rightarrow -\infty$. If these limits are attained
{\it uniformly} in $( x_1,x_2,\dots,x_p )$, then it is known  that for all $p$, all solutions of (\ref{eqn}) depend only on $x_{p+1}$.\footnote{This is known, for reasons that are only partially clear to G.W.G.~(one of the present authors), as the {\it Gibbons Conjecture} and has been proved by a number of people, but not by G.W.G.}

If we merely require that $\p\phi/\p x_{p+1}>0$, $\phi$ is bounded, and that
\ben
	V^\prime (\phi) = -\phi(1-\phi^2)\,,
\label{Allen}
\een
then it is known that for $p<8$, all solutions of (the stationary version of) this so-called Allen-Cahn equation~\cite{AC}
are planar.\footnote{This is known as
{\it de Giorgi's conjecture}~\cite{Giorgi}. He, for good reason, added the {\it caveat}
\lq\lq at least for $p<8$\rq\rq.}
However, if $p \ge 8$, then there are non-planar examples~\cite{Pino}.
In other words, the behavior of domain walls of finite thickness mirrors that of infinitesimally thin
domain walls.

Physically one expects that a stable minimal surface, such as a {\it catenoid} in ${\Bbb E}^3$,
could be mimicked by solution of the Allen-Cahn equation.
In fact, a numerical simulation showed that starting with a configuration for which $\phi=+1$ in the deep
interior of a catenoid  and $\phi=-1$ outside it, and allowing it to
relax to an energy minimizer does lead to a thick catenoidal domain wall~\cite{Sutcliffe}. More recently, G\`o\`zd\`z and Holyst~\cite{Gozdz} have constructed periodic minimal surfaces from Landau-Ginzburg models.

A powerful general argument is known that interfaces in media of a type should be either planar, spherical, or cylindrical. Serrin~\cite{Serrin} gave the proof of this fact for the media introduced by Koretweg~\cite{Korteweg}.
It is of interest to see how it breaks down in the present cases, {\it i.e.}, the minimal surface and the domain wall described by the Allen-Cahn equation. We discuss it in Appendix~\ref{sec:Korteweg}.

\section{Branes in Curved Backgrounds}
\label{sec:BH}

To obtain  a spacetime picture of the situation, we  
pass to the ultra static spacetime
$ M = {\Bbb R} \times \Sigma$ with spacetime metric
\ben
	ds^2 = - dt^2 + h\,.
\een  
The $p$-dimensional minimal surface  $S \subset \Sigma$ lifts  to a  
static solution of the equations of motion of a $(p+1)$-dimensional Lorentzian 
submanifold of $M$ governed by the Dirac-Nambu-Goto action. All such static solutions extremize the energy $E$ and will be stable if the energy is minimized. The energy $E$ in this case is just $p$-volume $ {\rm Vol}(S) $ with respect to the pull-back of the metric $h$ on $\Sigma$ to the minimal hypersurface $S$. Of course, we could also \lq \lq Wick rotate\rq \rq and set $t=-i\tau$ with $\tau$ real. We then have a $(p+1)$-dimensional submanifold of $M$ equipped with the Riemannian metric
\ben
	ds_E^2 = d \tau ^2 + h\,. 
\een
 
This construction becomes less trivial if the spacetime metric 
is static, but not ultra-static, {\it i.e.}, of the form
\ben
	ds^2 = - V^2(x) dt ^2 + h\,,
\een
where $x$ is the coordinates on $\Sigma$.
In that case the energy $E$ and the $p$-volume ${\rm Vol}(S)$ with respect to the pull-back 
$h_S$ of the metric $h$ on $\Sigma$ to the minimal hypersurface $S$ differ. 
In fact, if $\sigma$ is a local coordinate system on $S$ 
\ben
	E
	=
	\int_S V(x(\sigma)) \sqrt{h_S}\, d^p \sigma\,.  
\een
As before, we can Wick rotate and consider a $(p+1)$-minimal
hypersurface of the $(p+2)$-dimensional Riemannian manifold equipped with Riemannian metric
$ds_E^2 = V^2 d\tau^2 + h$.

\subsection{Cones near Black Holes}
\label{sec:Frolov}

A recent related example is that of Frolov~\cite{Frolov}, who considered a static $p$-brane in the $N$-dimensional Schwarzschild-Tangherlini black hole.\footnote{In~\cite{Frolov}, he considered a general case with $N \geq p+2$. Since the value of $N-p$ does not affect our argument, we only consider the hypersurface case, $N=p+2$. Note also that his argument is independent of the specific form of the background solution if it is spherically symmetric and has a non-degenerate horizon. See also~\cite{Frolov2} for the original works on a membrane in 4-dimensional black holes, and~\cite{FrolovGorbonos} for an interesting generalization to the case that the Dirac-Nambu-Goto action is corrected by quantum effects.} He showed that this gives a geodesic of the metric
\ben
	g
	=
	\left( r \sin \theta \right) ^{2p-2}
	\left[
		dr ^2
		+
		r^2 f(r) d \theta ^2
	\right]\,,
\;\;\;
	f(r) = 1 - \left( \frac{r_0}{r} \right)^{N-3}\,,
\;\;\;
	N = p+2\,.
\label{eq:frolov-g}
\een
Here, $r$ is a Schwarzschild radial coordinate and $\theta$ a co-latitude coordinate.
He found a geodesic of the metric (\ref{eq:frolov-g}) corresponding to a cone of which apex touches the horizon, and showed that a qualitatively different behavior sets in when the spacetime dimension $N \geq 8$ (or $p \geq 6$). On the face of it, this looks different from the result of Bombieri {\it et al.}~\cite{Bombieri}. However, as mentioned above, a static $p$-brane in an $N$-dimensional static Lorentzian manifold (with a periodic imaginary time) may be thought of as a ($p+1$)-brane in an $N$-dimensional Riemannian manifold. Thus, from the Riemannian point of view, the qualitatively different behavior happens when the submanifold has dimension 7 or larger. This agrees with what the analysis of minimal cones in Sec.~\ref{sec:flat} indicates.

To check the above observation, let us write the $N$-dimensional Schwarzschild-Tangherlini metric as
\bea
	ds^2
	=
	- f(r) dt^2 + f(r)^{-1} dr^2
    + r^2 ( d\theta^2 + \sin^2 \theta d\Omega_{N-3}^2 )\,.
\eea
Then, we focus on a local region near the south pole of horizon by setting
\bea
	r =r_0 + \xi\,,
\;\;\;
	\theta = \pi - \eta
\eea
with small $\xi/r_0$ and $\eta$. At the leading order, the metric is written as
\bea
	ds^2
	=
	- \frac{ (N-3) \xi }{ r_0 } dt^2
	+ \frac{ r_0 }{ (N-3) \xi } d\xi^2
	+ r_0^2 ( d\eta^2 + \eta^2 d\Omega_{N-3}^2 )\,.
\eea
Furthermore, introducing the following local coordinates
\bea
	x
	=
	\sqrt{ \frac{ 4r_0 \xi }{ N-3 } }\,,
\;\;\;\;
	y=
	r_0 \eta\,,
\eea
the near horizon metric reduces to
\bea
	ds^2
	=
	- \kappa^2 x^2 dt^2 + dx^2
	+
	dy^2 + y^2 d\Omega_{N-3}^2\,,
\label{eq:NH-metric}
\eea
where $\kappa = (N-3)/2r_0$ is the surface gravity.

Thus, the near-horizon effective 2-dimensional metric in which the geodesic is to be found is
\bea
	g = x^2 y^{2(N-3)} ( dx^2 + dy^2 )\,.
\label{eq:NH-g}
\eea
The problem has been reduced to that of $(m,n)=(1,N-3)$ in Sec.~\ref{sec:primary}.
Note that the factor $x^2$ in $g$ comes from the time component of metric~(\ref{eq:NH-metric}).
Thus, the cone $y=\sqrt{N-3}\;x$ is a geodesic near the horizon, and from the analysis of geodesic deviation, this geodesic corresponds to a minimizer if $ N = p+2 \geq 8 $.

The work by Frolov was in part motivated by that of Kol~\cite{Kol1} in which the ``merger transition'' from the Kaluza-Klein black holes to a black string was investigated. The black-hole--brane system indeed serves as a toy model of the merger transition and is shown to possess a critical dimension (we will review Kol's observation in Sec.~\ref{sec:discussion}). In addition, this system serves as the simplest (as far as we know) example of critical phenomena in gravitational systems~\cite{Choptuik}. The cone solution separates two phases of the brane: one has a Minkowski topology and another a black hole topology. The change of stability nature of the brane appears at $p=6$ and results in that of mass scaling of the black hole on the brane. It seems that the self similarity of the critical solution changes from discrete one to continuous one. It would be interesting to clarify why the breakdown of Bernstein conjecture is related to this change of self similarity in detail.

In addition, the black-hole--brane system has many applications to the physics of fundamental interactions via the AdS/CFT correspondence. The holographic dual of the phase transition from the Minkowski embedding to the brane embedding corresponds to the meson melting phase transition of matter in the fundamental representation (see, {\it e.g.}, Refs.~\cite{Mateos1}). Although the systems investigated in the literature so far correspond to the black-hole--brane systems below the critical dimension (as far as we know), it would be interesting to see in what the failure of the Bernstein conjecture results in the gauge theory side.   


\subsection{Truncated Cone in Black Hole: An Exact Solution}
\label{sec:truncated}

The minimal cone in Sec.~\ref{sec:Frolov} is the example of near-horizon approximate solution, which deviates from a cone away from the horizon. In this subsection, we present an example of exact solution that is minimal and ``globally straight''.

Any static $N$-dimensional spacetime metric invariant under $\mathrm{SO}(N-1)$ can be written
in isotropic, not Schwarzschild, coordinates as
\ben
	ds^2
	=
	-A^2(\hat{r}) dt^2 + B^2 (\hat{r}) \bigl
	(  d\hat{r}^2 + \hat{r}^2 d\Omega^2_{N-2} \bigr ) \,.
\label{eq:N-Sch1}
\een 
The metric in parentheses is the standard metric on ${\Bbb E} ^{N-1}$, and $\mathrm{SO}(N-1)$ and its subgroups
act in the standard way. Thus, we can apply the equivalent variational
methods described in~\cite{Fomenko,DaoFomenko}. The introduction 
of the coordinates works as before:
\bea
	ds^2
	=
	-A^2(\hat{r}) dt^2 + B^2 (\hat{r})
	\left(
		dx^2 + x^2 d\Omega_m^2
		+
		dy^2 + y^2 d\Omega_n^2
	\right)\,,
\eea
where
\bea
	\hat{r}^2 = x^2 + y^2\,,
\;\;\;\;\;
	N = m+n+3\,,
\;\;\;\;\;
	m,n \geq 1\,.
\eea
We assume that a hypersurface configuration is static and invariant under the action of
$\mathrm{SO}(m+1) \times \mathrm{SO}(n+1)$ with the the standard action on $\mathbb{E}^{m+n+2} = \mathbb{E}^{m+1} \times \mathbb{E}^{n+1}$.
Then, the hypersurface may be given by $y=y(x)$, and the problem to find the minimal surface is reduced to find the geodesic $\gamma$ of the 2-dimensional space whose metric is given by
\bea
	g
	=
	d\ell^2
	=
	A^2 B^{ 2(m+n+1) } x^{2m} y^{2n} ( dx^2 + dy^2 )\,.
\label{eq:g-BH}
\eea
The geodesic equation is given by
\bea
	- \frac{ y^{\prime\prime} }{ ( 1+y^{\prime 2} )^{3/2} }
	+
	\frac{ nx - myy^\prime }{ xy \sqrt{ 1+y^{ \prime 2 } } }
	+
	\frac{ ( y-xy^\prime ) [ A^\prime B + (m+n+1)AB^\prime ] }{ \hat{r} AB \sqrt{ 1+y^{\prime 2} } }
	=
	0\,,
\label{eq:EL-BH}
\eea
where the prime denotes the differentiation with respect to each argument ({\it i.e.}, $y^\prime = \pd_x y$, $A^\prime = \pd_{\hat{r}} A$, and so on).
The last term represents the effect of curvature of background geometry.
Note that a horizon, where $A=0$, is a singular point of this equation.
One can show the following {\it truncated cone} solves Eq.~(\ref{eq:EL-BH}),
\bea
	y = \sqrt{ \frac{n}{m} }\, x\,,
\;\;\;\;\;\;
	x \geq \sqrt{ \frac{ m }{ m+n } }\, \hat{r}_0\,.
\label{eq:tcone}
\eea
Here, $\hat{r}_0$ is the location of outermost horizon in the isotropic coordinates.
This solution is an exact solution unlike the approximate cone in Sec.~\ref{sec:Frolov}.

One can examine whether the cone~(\ref{eq:tcone}) is a minimizer or not
with the Jacobi equation.
The cone (\ref{eq:tcone}) is parameterized by the isotropic coordinate $\hat{r}$ as
\bea
	x
	=
	\sqrt{ \frac{ m }{ m+n } }\; \hat{r}\,,
\;\;\;
	y
	=
	\sqrt{ \frac{ n }{ m+n } }\; \hat{r}\,.
\eea
Thus, the relation between $\hat{r}$ and the proper length $\ell$ along the cone is
\bea
	\frac{ d\ell }{ d\hat{r} }
	=
	\frac{ 1 }{ G(\hat{r}) }
	:=
	\frac{ m^{m/2} n^{n/2} }{ ( m+n )^{ (m+n)/2 } } 
	A B^{m+n+1} \hat{r}^{ m+n }\,.
\eea
Changing variable from $\ell$ to $\hat{r}$, the Jacobi equation (\ref{eq:gd}) is written as
\bea
	G^2 \frac{ d^2 \eta }{ d\hat{r}^2 }
	+
	G \frac{ d G }{ d \hat{r} } \frac{ d \eta }{ d \hat{r} }
	+
	K \eta = 0\,,
\eea
where $K$ for metric~(\ref{eq:g-BH}) is given by
\bea
&&
	K
	=
	\frac{ 1 }{ A^2 B^{2(m+n+1)} x^{2m} y^{2n} }
	\left( \frac{m}{x^2} + \frac{n}{y^2} \right)
\nonumber
\\
&&
	-
	\frac{ 1 }{ \hat{r} A^4 B^{ 2(m+n+2) } x^{2m} y^{2n} }
	\left[
		( AA^\prime - \hat{r}A^{\prime 2} + \hat{r} AA^{\prime\prime} ) B^2
		+
		( m+n+1 ) A^2 ( BB^\prime - \hat{r}B^{\prime 2} + \hat{r}BB^{\prime\prime} )
	\right]\,.
\nonumber
\\
\eea
Here, the prime denotes the differentiation with respect to $\hat{r}$.

For simplicity, let us assume $N=(m+n+3)$-dimensional Schwarzschild-Tangherlini black hole as the background, which is given with the standard Schwarzschild coordinate $r$ by
\bea
	ds^2
	=
	-f(r) dt^2 + f(r)^{-1} dr^2 + r^2 d\Omega_{N-2}^2\,,
\;\;\;
	f = 1 - \left( \frac{r_0}{r} \right)^{m+n}\,.
\label{eq:N-Sch2}
\eea
The coordinate transformation between $\hat{r}$ and $r$ [see Eqs.~(\ref{eq:N-Sch1}) and (\ref{eq:N-Sch2})] are
\bea
	A
	=
	f^{1/2}\,,
\;\;\;
	B
	=
	\frac{r}{\hat{r}}\,,
\;\;\;
	\frac{ dr }{ d\hat{r} }
	=
	\frac{ rf^{1/2} }{ \hat{r} }\,.
\eea
After some calculations we have the Gauss curvature in the Schwarzschild coordinates,
\bea
	K
	=
	\frac{ (m+n)^{m+n +1} }{ 4m^m n^n }
	\cdot
	\frac{ 8r^{ 2(m+n) } - 10r^{m+n} + (m+n +2) }{ r^{ 2( m+n +1 ) } ( r^{m+n} -1 )^2 }\,,
\label{eq:Gauss-Sch2}
\eea
where we have set $r_0=1$. One can see that from Eq.~(\ref{eq:Gauss-Sch2}) the Gauss curvature is positive definite for the present case $m+n \geq 2$ and divergent at the horizon $r=1$.
Thus, one gets the Jacobi equation
\bea
	\frac{d^2 \eta}{ dr^2 }
	-
	\frac{m+n}{r} \frac{d \eta}{ dr }
	+
	\frac{ (m+n) [ 8r^{2(m+n)} - 10r^{m+n} + m+n+2 ] }{ 4 r^2 (r^{m+n}-1)^2 } \eta
	=
	0\,.
\label{eq:Jacobi-Sch}
\eea
Now, we can see that the instability exists at large distance by considering the asymptotic region.\footnote{
If we put $\eta = r^{(m+n)/2}\tilde{\eta} $, the first derivative term vanishes,
\bea
	\frac{ d^2 \tilde{\eta} }{ dr^2 }
	+
	\frac{ (m+n) r^{m+n} [ - (m+n-6) r^{m+n} + 2( m+n-3 ) ] }
		 { 4 r^2 ( r^{m+n} - 1 )^2 } \tilde{\eta}
	= 0\,.
\label{eq:deviation-Sch3}
\eea
Furthermore, if one introduces a new variable $X=r^{m+n}-1$, Eq.~(\ref{eq:deviation-Sch3}) becomes
\bea
	\frac{d^2 \tilde{\eta}}{ dX^2 }
	+
	\frac{ m+n-1 }{ ( m+n )( X+1 ) } \frac{d \tilde{\eta}}{ dX }
	+
	\frac{ - (m+n-6) X + m + n }{ 4 ( m+n ) X^2 ( X+1 )  } \tilde{\eta}
	=
	0\,.
\label{eq:deviation-Sch4}
\eea
Analytic solutions to Eqs.~(\ref{eq:deviation-Sch3}) and (\ref{eq:deviation-Sch4}) can be obtained explicitly in terms of special functions. However, they are not so informative and therefore omitted to be written down here.
}
For $r\gg 1$, Eq.~(\ref{eq:Jacobi-Sch}) reduces to
\bea
	\frac{d^2 \eta}{ dr^2 }
	-
	\frac{m+n}{r} \frac{d \eta}{ dr }
	+
	\frac{ 2(m+n)}{ r^2 } \eta
	=
	0\,
\eea
and is solved by
\bea
	\eta = r^{\beta_\pm}\,,
\;\;\;\;
	\beta_\pm
	=
	\frac12
	\left(
		m+n+1 \pm \sqrt{ (m+n)^2 - 6 (m+n) + 1 }
	\right)\,.
\eea
Thus, the geodesic deviation oscillates for $5 \leq N = m+n+3 \leq 8$, while not for $ N \geq 9$ implying the cone is a minimizer.

\section{Discussion}
\label{sec:discussion}

In this last section, we review some examples of physical system in the literature whose behavior can be related to the Bernstein conjecture and its breakdown. These examples suggest that the consequences of Bernstein conjecture and its breakdown in general dimensions appear in a variety of systems either in an explicit or non-explicit way. It would be interesting to investigate in detail what to extent the behaviors of these physical systems are related to those seen above.


\paragraph{Cone as Einstein Manifold.}

We have seen so far that there exist critical dimensions for the minimal cones,
which provide some interesting consequences if we suppose the brane theory point of view as mentioned in Introduction, or if we consider the black-hole--brane systems. 
We will see here, however, that there exists a critical dimension also for a cone that is an Einstein manifold. This suggests that the stability of spacetimes that have the cone as a part of them changes at a certain dimension. As an example, let us see the Kol's observation on a {\it Ricci flat cone}~\cite{Kol1} (see also Sec.~VI in~\cite{Gibbons:2002th}). He considered the cone over $S^2 \times S^{2}$ and its generalizations to $S^m \times S^n$ in the modeling of (Euclidean version of) ``merger transition'' from caged black holes to a black string. The Einstein equations [{\it i.e.}, (Ricci tensor)=0] for metric
\bea
	ds_E^2
	=
	dx^2 + e^{2a(x)} d\Omega_m^2 + e^{2b(x)} d\Omega_n^2
\eea
are given by
\bea
&&
	a^{\prime\prime} 
	+ ma^{\prime 2} + n a^{\prime} b^\prime - (m-1) e^{-2a} = 0\,,
\;\;\;\;
	b^{\prime\prime} + nb^{\prime 2} + ma^{\prime} b^\prime -(n-1) e^{-2b} = 0\,,
\nonumber
\\
&&
	m(m-1) a^{\prime 2} + 2mn a^{\prime} b^{\prime} + n(n-1) b^{\prime 2}
	-
	m(m-1) e^{-2a} - n(n-1)e^{-2b} = 0\,.
\eea
One can check the following Ricci flat cone solves the above equations,
\bea
	ds_E^2
	&=&
	dx^2 + e^{a_0(x)} d\Omega_m^2 + e^{b_0(x)} d\Omega_n^2
\nonumber
\\
	&:=&
	dx^2 + \frac{m-1}{m+n-1} x^2 d\Omega_m^2 + \frac{n-1}{m+n-1} x^2 d\Omega_n^2\,.
\label{eq:flat-cone}
\eea
Then, he considered the perturbation around solution~(\ref{eq:flat-cone}) by setting
$a = a_0 + a_1(x)$, $b = b_0 + b_1(x)$, and linearizing the Einstein equations with respect to $a_1$ and $b_1$. The combinations $ a_+ := m a_1 + n a_2$ and $a_- := a_1 - a_2$ decouple the perturbation equations. The equation for the gauge invariant combination, $a_-$, is 
\bea
	a_-^{\prime\prime} + \frac{ m+n }{ x }a_-^\prime + \frac{ 2(m+n-1) }{ x^2 }a_- = 0\,.
\eea
This equation is solved by
\bea
	a_- 
	=
	x^{\beta_\pm}\,,
\;\;\;\;\;
	\beta_\pm
	=
	-\frac12 \left( m+n-1 \pm \sqrt{ (m+n-9)(m+n-1) } \right)\,.
\eea
This mode corresponds to the shrinking of one sphere and expanding of the other. The discriminant is non-negative if $m+n \ge 9$, which implies the Ricci flat cone is stable for $m+n+1 \geq 10$, (precisely speaking, it is marginally stable for $m+n+1=10$). 

The above observation on the cone over $S^m \times S^n$ might 
be related to a change of stability for the spacetimes containing the Ricci flat cone as their internal space. Indeed, in~\cite{DeWolfe:2001nz} $\mathrm{AdS}_q \times S^{m} \times S^{n}$ was shown to be unstable for $m+n<9$ due to the violation of Breitenlohner-Freedman mass bound in AdS~\cite{Breitenlohner:1982bm}, while it was shown to be stable for $m+n \geq 9$.

\paragraph{Non-Zero-Constant Mean Curvature Surfaces.}

A minimal surface is characterized by the vanishing of its mean curvature,
which is a mathematical model of soap film containing no air in it.
When one considers the soap bubble containing air, the configuration has
a non-zero-constant mean curvature. Recently, the bifurcation structures of axially symmetric constant mean curvature surfaces in general dimensions were revealed by two of the present authors~\cite{MiyamotoMaeda}, which was motivated by the existence of critical dimensions in the black-hole--black-string system~\cite{Sorkin}. The key observation that can be related to the failure of Bernstein conjecture is that there appears a new branch of undulating cylinder appears at 9 space dimension. This branch was shown to be stable in~\cite{Miyamoto} with the so-called surface diffusion equation~\cite{Mullins}, which is closely related to the (original dynamical version of) Allen-Cahn equation.

The recent gravity/fluid correspondence predicts that a sort of black holes localized in the IR of AdS are dual to the fluid lumps whose surfaces have constant mean curvatures~\cite{Aharony}. In such a theory, the minimal surfaces and constant mean curvature surfaces have rather physical meanings via the correspondence. Thus, it would be interesting to investigate their relations to the breakdown of Bernstein conjecture in more detail.

\subsection*{Acknowledgments}


We would like to thank Costas Bachas, Valeri P.~Frolov, Dan Gorbonos, Barak Kol, and Oleg Lunin for helpful discussions and comments. K.M.~and U.M.~would like to acknowledge the hospitality during their stay  (Sep.~2008) in DAMTP and the Centre for Theoretical Cosmology, Cambridge University,  where this work was started. K.M.~and U.M.~acknowledge also the organizers of workshop ``Black Holes VII: Theory and Mathematical Aspects'' (May 2009) at Banff, Alberta, Canada. K.M.~is supported in part by the Grant-in-Aid for Scientific Research Fund of the JSPS (No.~19540308) and for the Japan-U.K. Research Cooperative Program, and by the Waseda University Grants for the Special Research Projects. U.M.~is supported by the Golda Meir Fellowship, the Israel Science Foundation Grant (No.~607/05), and the DIP Grant (No.~H.52).

\appendix

\section{Korteweg's Theory of Phase Equilibrium}
\label{sec:Korteweg}

Serrin's version~\cite{Serrin} of Korteweg's theory~\cite{Korteweg} starts with the equilibrium condition for the spatial stress tensor
\ben
	\partial_i T_{ij} =0\,,
\;\;\;\;
	(i,j=1,2,\ldots,p+1)\,,
\label{stress}
\een
where $T_{ij}$ is given in terms of a density function $\rho(x)$ by  
\bea
&&
	T_{ij}
	=
	- P \delta_{ij}
	+
	\left(
		\alpha \nabla ^2 \rho
		+
		\beta  | \nabla \rho |^2
	\right) \delta_{ij} 
	+
	\left(
		\gamma \p_i \p_j \rho
		+
		\delta \p_i \rho \p_j \rho
	\right)\,,
\eea
where $|\nabla \rho| := ( \delta^{ij} \partial_i \rho \partial_j \rho )^{1/2}$ and
$(P,\alpha,\beta,\gamma,\delta)$ is a set of functions of $\rho$.
Starting from these equations, Serrin derived the over-determined system of equations,
\ben
	\nabla ^2 \rho
	=
	\xi(\rho) \,,\qquad
	\bigl| \nabla \rho \bigr|
	=
	\zeta(\rho)\,,
\label{over} 
\een
where $\xi$ and $\zeta$ are some functions of $\rho$ given in terms of $(P,\alpha,\beta,\gamma,\delta)$. It was then shown by Pucci~\cite{Pucci} that solutions $\rho(x)$ of Eq.~(\ref{over})
must have level sets given either by concentric spheres, cylinders or parallel planes.

Now, let us see the above Serrin's argument in some detail. If the following set of functions are defined
\ben
	a
	:=
	\alpha + \gamma\,,
\qquad
	b
	:=
	\beta +\delta\,,
\qquad
	c
	:=
	\gamma ^\prime -\delta\,,
\qquad
	( \; {}^\prime := \pd_\rho )\,,
\een
Eq.~(\ref{stress}) implies that
\ben
	\p_i
	\left[
		-P + a \nabla ^2 \rho +
		\left(
			b + \half c
		\right) |\nabla \rho |^2
	\right]
	=
	\Bigr(
		c \nabla ^2 \rho + \half c^\prime |\nabla \rho |^2
	\Bigl) \p_i \rho\,.  
\een
Now if 
\ben
	A := bc + \half (c^2 - a c^\prime ) \ne 0\,,
\een
then he claims to be able to establish that Eq.~(\ref{over}) holds for an
appropriate choice of $\xi(\rho)$ and $\zeta(\rho)$. To this end, he defines
\ben
	F
	:=
	-P + a \nabla ^2 \rho +
	\left(
		b + \half c
	\right) |\nabla \rho |^2\,, \qquad
	G
	:=
	c \nabla ^2 \rho + \half c^\prime |\nabla \rho |^2\,,
\een
so that $\p_i F = G \p_i \rho$.
Thus, there exists a real valued function $\omega(\rho)$ such that $F = \omega(\rho)$ and $G = \omega^\prime (\rho)$, and hence
\bea
&&
	\nabla ^2 \rho
	=
	\xi(\rho)
	:=
	{1 \over A}
	\left[
		\left( b + \half c \right) \omega^\prime - c^\prime ( \omega+P )
	\right]\,,
\nonumber
\\
&&
	| \nabla \rho |^2
	=
	\zeta^2(\rho)
	:=
	{1 \over A}
	\left[
		c( \omega+P ) - a \omega^\prime
	\right]\,.
\eea

At first glance, the above Serrin's model is rather general,
and the Allen-Cahn domain walls and the minimal surfaces seem to be 
covered by  the  argument above. We will see below, however, 
the domain wall and minimal surface are exceptional cases for which his argument breaks down.

Firstly, let us consider a single scalar field $\phi$ whose spatial stress tensor is given by
\ben
	T_{ij}^{(\phi)}
	=
	\p_i\phi \p_j\phi - \half \delta_{ij} \left[ ( \p_k \phi )^2 + 2 V(\phi)  \right]\,.
\een
Its divergence is given by
\ben	
	\p_i T_{ij}^{(\phi)}
	=
	\p_j \left[ \Delta \phi - V^\prime (\phi) \right]\,.
\een
Thus, we just get Eq.~(\ref{eqn}).
In Serrin's notation, taking $\rho=1$, we have $P=-V$, $ (\alpha,\beta,\gamma,\delta ) = (0,-1/2, 0,1)$
whence $(a,b,c) = (0,1/2,-1)$.
Thus, we have
\ben
	F = V\,,
\qquad
	G = \Delta \phi\,,
\qquad
	A = 0\,,
\een
which is the case excluded by Serrin.

Secondly, let us consider the minimal surfaces. The non-parametric form of the minimal 
surface equation [denote the minimal surface by $\varphi(x)=0$] can be derived by extremizing the energy functional
\ben
	E[\varphi]
	=
	\int \mathcal{E}(\varphi, \p_i \varphi) d^{p+1} x
	=
	\int  \Bigl ( \sqrt{ 1 + |\nabla \varphi|^2 } - 1 \Bigl ) \, d^{p+1} x\,.
\een
Thus, the stress tensor is given by
\ben
	T_{ij}^{(\varphi)}
	=
	{
	\p_i \varphi \p_j \varphi
	\over
	\sqrt{1+ |\nabla \varphi|^2 }
	}   
	-
	\delta_{ij} \Bigl( \sqrt{ 1+ |\nabla \varphi|^2 } - 1 \Bigl)\,.  
\een
This is not of the form introduced by Korteweg, and Serrin's argument does not work for the minimal surfaces.
Moreover, one has
\ben
	\p_i T_{ij}^{(\varphi)}
	=
	\p_j \varphi \;
	\p_i
	\left(
		{ \p_i \varphi \over   \sqrt{1+ |\nabla \varphi|^2 }   }
	\right)
	\equiv 0
\een
by virtue of the equation of motion.
This will always be true for the systems obtained by varying an energy
functional, $\int \mathcal{E}(\varphi,\p_i\varphi) d^{p+1} x$.



\begin{thebibliography}{99}

\bibitem{Bombieri}
E.~Bombieri, E.~de Giorgi and E.~Giusti,
``Minimal Cones and the Bernstein Problem'',
{\it Inventiones math.} {\bf 7}, 243-268 (1969).

\bibitem{Schoen1}
  R.~Schoen and S.~T.~Yau,
  ``Positivity Of The Total Mass Of A General Space-Time,''
  Phys.\ Rev.\ Lett.\  {\bf 43}, 1457 (1979);
  R.~Schoen and S.~T.~Yau,
  ``On the Proof of the Positivity Mass Conjecture in General Relativity,''
  Comm.\ Math.\ Phys.\  {\bf 65}, 45 (1979).

\bibitem{Bray}
H.~L.~Bray and D.~A.~Lee,
``On the Riemannian Penrose inequality in dimensions less than 8'',
arXiv:0705.1128 [math.DG].

\bibitem{Schwartz:2007gj}
  F.~Schwartz,
  ``Existence of outermost apparent horizons with product of spheres
  topology,''
  arXiv:0704.2403 [gr-qc].

\bibitem{Frolov}
  V.~P.~Frolov,
  ``Merger transitions in brane-black-hole systems: Criticality, scaling,  and
  self-similarity,''
  Phys.\ Rev.\  D {\bf 74}, 044006 (2006)
  [arXiv:gr-qc/0604114].

\bibitem{Fomenko}
A.~T.~Fomenko,
``The Plateau Problem: Part II the present state of the theory'',
Studies in the Development of Modern Mathematics Volume 1,
Gordon and Breach Science Publishers (1990).

\bibitem{Rauch}
H.~E.~Rauch,
``Geodesic and Curvature in Differential Geometry in the Large'',
Graduate School of Mathematical Sciences,
Yeshiva University, New York (1959);
M.~M.~ Postnikov,
``The Variational Theory of Geodesics'',
Dover Publications (1983).  

\bibitem{DaoFomenko}
Dao Trong Thi and A.~T.~Fomenko,
``Minimal surfaces, stratified multivarifolds, and the plateau problem'',
Translations of mathematical monographs, AMS, {\bf 84} (1991).

\bibitem{AC}
S.~Allen and J.~W.~Cahn,
``A microscopic theory for antiphase boundary motion and its application to antiphase domain coarsening'',
Acta Metall., {\bf 27}, 1084-1095 (1979).

\bibitem{Pino}
M.~del Pino, M.~Kowalczyk and J.~Wei,
``On de Giorgi conjecture in dimension $N\geq 9$'',
arXiv:0806.3141 [math.AP].

\bibitem{Giorgi}
E.~de~Giorgi,
``Convergence problems for functionals and operators'',
Proc.~Int.~Meeting on Recent Methods in Nonlinear Analysis (Rome, 1978),
131-188, Pitagora, Bologna (1979).

\bibitem{Sutcliffe}
Paul Sutcliffe,
private communication.

\bibitem{Gozdz}
W.~Gozdz and R.~Holyst,
``From the Plateau problem to periodic minimal surfaces in lipids, surfactants and diblock copolymers'',
cond-mat/9604003;
``High Genus Periodic Gyroid Surfaces of Nonpositive Gaussian Curvature''
Phys.\ Rev.\ Lett. {\bf 76}, 2726 (1996).

\bibitem{Serrin}
J.~Serrin,
``The form of interfacial surfaces in Korteweg's theory of phase equilibria'',
Quart.\ Appl.\ Math. {\bf 41}, No.~3, 357-364 (1983/84).

\bibitem{Korteweg}
D.~J.~Korteweg,
``Sur la forme que prennat les \'equations du mouvement des fluides si l'on tient
compte des forces capilaires caus\'ees par des variations de density'',
Archives N\`eerlandaise des Science Exactes et Naturelles {\bf 6}, 1-24 (1901). 

\bibitem{Frolov2}
  M.~Christensen, V.~P.~Frolov and A.~L.~Larsen,
  ``Soap bubbles in outer space: Interaction of a domain wall with a black
  hole,''
  Phys.\ Rev.\  D {\bf 58}, 085008 (1998)
  [arXiv:hep-th/9803158];
  V.~P.~Frolov, A.~L.~Larsen and M.~Christensen,
  ``Domain wall interacting with a black hole: A new example of critical
  phenomena,''
  Phys.\ Rev.\  D {\bf 59}, 125008 (1999)
  [arXiv:hep-th/9811148].

\bibitem{FrolovGorbonos}
  V.~P.~Frolov and D.~Gorbonos,
  ``A Toy Model for Topology Change Transitions: Role of Curvature
  Corrections,''
  arXiv:0808.3024 [hep-th].
  
\bibitem{Kol1}
  B.~Kol,
  ``Topology change in general relativity and the black-hole black-string
  transition,''
  JHEP {\bf 0510}, 049 (2005)
  [arXiv:hep-th/0206220].

\bibitem{Choptuik}
  M.~W.~Choptuik,
  ``Universality And Scaling In Gravitational Collapse Of A Massless Scalar
  Field,''
  Phys.\ Rev.\ Lett.\  {\bf 70}, 9 (1993).

\bibitem{Mateos1}
  D.~Mateos, R.~C.~Myers and R.~M.~Thomson,
  ``Holographic phase transitions with fundamental matter,''
  Phys.\ Rev.\ Lett.\  {\bf 97}, 091601 (2006)
  [arXiv:hep-th/0605046];
  D.~Mateos, R.~C.~Myers and R.~M.~Thomson,
  ``Thermodynamics of the brane,''
  JHEP {\bf 0705}, 067 (2007)
  [arXiv:hep-th/0701132];
  S.~Kobayashi, D.~Mateos, S.~Matsuura, R.~C.~Myers and R.~M.~Thomson,
  ``Holographic phase transitions at finite baryon density,''
  JHEP {\bf 0702}, 016 (2007)
  [arXiv:hep-th/0611099].




\bibitem{Gibbons:2002th}
  G.~W.~Gibbons, S.~A.~Hartnoll and C.~N.~Pope,
  ``Bohm and Einstein-Sasaki metrics, black holes and cosmological event
  horizons,''
  Phys.\ Rev.\  D {\bf 67}, 084024 (2003)
  [arXiv:hep-th/0208031].

\bibitem{DeWolfe:2001nz}
  O.~DeWolfe, D.~Z.~Freedman, S.~S.~Gubser, G.~T.~Horowitz and I.~Mitra,
  ``Stability of AdS(p) x M(q) compactifications without supersymmetry,''
  Phys.\ Rev.\  D {\bf 65}, 064033 (2002)
  [arXiv:hep-th/0105047];
  T.~Shiromizu, D.~Ida, H.~Ochiai and T.~Torii,
  ``Stability of AdS(p) x S(n) x S(q-n) compactifications,''
  Phys.\ Rev.\  D {\bf 64}, 084025 (2001)
  [arXiv:hep-th/0106265].

\bibitem{Breitenlohner:1982bm}
  P.~Breitenlohner and D.~Z.~Freedman,
  ``Positive Energy In Anti-De Sitter Backgrounds And Gauged Extended
  Supergravity,''
  Phys.\ Lett.\  B {\bf 115}, 197 (1982).

\bibitem{MiyamotoMaeda}
  U.~Miyamoto and K.~i.~Maeda,
  ``Liquid bridges and black strings in higher dimensions,''
  Phys.\ Lett.\  B {\bf 664}, 103 (2008)
  [arXiv:0803.3037 [hep-th]].

\bibitem{Sorkin}
  E.~Sorkin,
  ``A critical dimension in the black-string phase transition,''
  Phys.\ Rev.\ Lett.\  {\bf 93}, 031601 (2004)
  [arXiv:hep-th/0402216];
  H.~Kudoh and U.~Miyamoto,
  ``On non-uniform smeared black branes,''
  Class.\ Quant.\ Grav.\  {\bf 22}, 3853 (2005)
  [arXiv:hep-th/0506019].
  
\bibitem{Miyamoto}
  U.~Miyamoto,
  ``Curvature driven diffusion, Rayleigh-Plateau, and Gregory-Laflamme,''
  Phys.\ Rev.\  D {\bf 78}, 026001 (2008)
  [arXiv:0804.1723 [hep-th]].
  
\bibitem{Mullins}
W.~W.~Mullins,
``Theory of Thermal Grooving'',
J.~Appl.~Phys.~{\bf 28} (3) (1957) 333.

\bibitem{Aharony}
  O.~Aharony, S.~Minwalla and T.~Wiseman,
  ``Plasma-balls in large N gauge theories and localized black holes,''
  Class.\ Quant.\ Grav.\  {\bf 23}, 2171 (2006)
  [arXiv:hep-th/0507219];
  K.~i.~Maeda and U.~Miyamoto,
  ``Black hole-black string phase transitions from hydrodynamics,''
  JHEP {\bf 0903}, 066 (2009)
  [arXiv:0811.2305 [hep-th]];
  M.~M.~Caldarelli, O.~J.~C.~Dias, R.~Emparan and D.~Klemm,
  ``Black Holes as Lumps of Fluid,''
  arXiv:0811.2381 [hep-th].
 
\bibitem{Pucci}
Pucci,
``Patrizia An overdetermined system'',
Quart.\ Appl.\ Math.\ {\bf 41}, No.~3, 365-367 (1983/84).

\end{thebibliography}
\end{document}